\documentclass[12pt]{amsart}
\usepackage{geometry} 
\usepackage[utf8]{inputenc} 
\usepackage{textcomp} 
\usepackage{graphicx}  
\usepackage{flafter}  

\usepackage{amsmath,amssymb}  
\usepackage{bm}  
\usepackage{memhfixc}  
\usepackage{pdfsync}  

\usepackage{amssymb,amsfonts,amsmath}

\usepackage{caption}

\title{An agent-based epidemiological model of incarceration}
\author{Kristian Lum$^1$, Samarth Swarup$^1$, Stephen Eubank$^1$, James Hawdon$^2$}
\begin{document}


\maketitle

\begin{center}
$^1$ Network Dynamics and Simulation Science Laboratory, Virginia Bioinformatics Institute, Virginia Tech\\
$^2$ Department of Sociology, Virginia Tech
\end{center}


\begin{abstract} 
  We build an agent-based model of incarceration based on the SIS model of infectious disease propagation. Our central hypothesis is that the observed racial disparities in incarceration rates between Black and White Americans can be explained as the result of differential sentencing between the two demographic groups. We demonstrate that if incarceration can be spread through a social influence network, then even relatively small differences in sentencing can result in the large disparities in incarceration rates. Controlling for effects of transmissibility, susceptibility, and influence network structure, our model reproduces the observed  large  disparities  in incarceration rates given the  differences in sentence lengths for White and Black drug offenders in the United States without extensive parameter tuning. We further establish the suitability of the SIS model as applied to incarceration, as the observed structural patterns of recidivism are an emergent property of the model.   In fact, our model shows a remarkably close correspondence with California incarceration data, without requiring any parameter tuning. This work advances efforts to combine the theories and methods of epidemiology and criminology.

\end{abstract}




\section{Introduction}
The rapid increase in the U.S. incarceration rate over the last few decades has been described as an epidemic. According to Bureau of Justice Statistics, the per capita rate of incarceration nearly quadrupled between 1978 and 2011, from 137 to 511 persons per 100,000  \cite{Carson:2013fk}. This prison boom has primarily affected Black Americans, especially Black males. By 2011, Black incarceration rates were over six times higher than White rates (3,023 per 100,000 for Blacks, and 478 per 100,000 for Whites). 

Racial disparities in incarceration rates have been studied extensively\cite{Oliver:2001uq, tonry2010social, mitchell2005meta, mauer2006race, wakefield2010incarceration}, and while these disparities are partially due to differences in criminal involvement \cite{pastorekathleen}, the increase in imprisonment for Black males since 1980 was not matched by a similar increase in Black-male criminality \cite{beckett2000politics, beckett2006race, western2006punishment}. What then accounts for the racial disparities in incarceration?  Scholars offer several explanations, including differential exposure to police surveillance \cite{beckett2006race}, prosecutorial discrimination \cite{rehavi2012racial}, the use of incarceration to deal with a ``racial threat'' \cite{bridges1988law, wang2010multilevel}, or sentencing disparities between Blacks and Whites. Although studies reveal that racial sentencing disparities are reduced when legal factors \cite{wooldredge2011victim, nicosia2013disparities, brennan2008race} or social contexts \cite{helms2010modeling, wooldredge2007convicting} are considered, a recent meta-analysis reports that sentencing disparities remain {\it even after controlling for these factors} \cite{mitchell2005meta}. While the magnitude of the difference is small and variable, it is largest in cases involving discretionary powers and for drug offenses.  In fact, it has been shown that Blacks receive longer sentences than Whites for drug offenses \cite{beckett2006race, tonry2010social}.  

Careful study of patterns of incarceration reveals that incarceration behaves like a contagious disease in that the close associates of an incarcerated person have higher-than-average probabilities of being incarcerated.  An individual's incarceration can be ``transmitted'' to others via several mechanisms.
It can increase the strains family members must endure, expose them to criminal norms, and enmesh them in a criminal subculture, thereby increasing the probability these people would themselves commit crimes \cite{agnew1992empirical, sutherland1974criminology} and be incarcerated. 
Alternatively, 
once a person is incarcerated, the police and courts pay more attention to the inmate's family and friends, thereby increasing the probability they will be caught, prosecuted and imprisoned \cite{west1973becomes, farrington2001predicting, besemer2011relationship}. 
Regardless of the mechanisms involved, the incarceration of one family member undoubtedly increases the likelihood of other family members being incarcerated \cite{wildeman2010paternal, wakefield2011mass, wildeman2009parental,thornberry2009apple}. This suggests that models of contagion may aptly characterize incarceration. 

It is well known that some models of contagion exhibit nonlinearities.  Nonlinear processes such as infectious disease outbreaks are capable of amplifying small differences in parameters through feedback in certain circumstances. One such example is the SIS (Susceptible-Infected-Suspectible) in which individuals transition between susceptible and infectious states-- near a critical value of transmission probability, positive feedbacks amplify small differences in transmission rate to create large differences in {\it prevalence}, the number of infected people.  

Here we explore the possibility that the dramatic racial disparity in incarceration rates is a consequence of a disparity in sentencing via analogous nonlinearities to those described. Our hypothesis is that the differences in sentencing between Blacks and Whites result in disparate transmission probabilities near a critical point, causing the incarceration epidemic to reach very different levels of prevalence under the two sentencing schemes. Using an agent-based SIS model, we simulate the spread of incarceration through a highly realistic synthetic population. We run our simulations under one scenario in which sentence lengths are consistent with those received by Black Americans for drug possession and under a second scenario in which the sentence lengths are representative of those received by Whites. 
We demonstrate here that if incarceration is infectious-like in that one's incarceration causes his or her family and friends to themselves become more likely than the general population to be incarcerated, then it is plausible that the observed large disparities in incarceration rates between Black and White Americans are the result of inequity in sentencing.

\section{The susceptible - incarcerated -susceptible (SIS) model}

The hypothesized ``transmissibility" of incarceration suggests that an SIS model in which incarceration is modeled as though it were an infectious disease is appropriate.  An agent-based simulation of the SIS model requires three main components: a  contact network through which individuals stochastically transmit the disease, transmission probabilities that dictate the rate at which agents transmit to each other, and a period of infectivity \cite{brauer2013mathematical}. Network ties represent opportunities for transmission, and the structure of the network through which the disease spreads affects the dynamics of the outbreak. In a disease model, network ties between agents typically denote physical contact or close proximity; in the case of incarceration, they denote the existence of a familial relationship or close friendship, i.e. the existence of a strong influence between the agents. 

 In the disease modeling paradigm, transmission probabilities may be a function of characteristics of the individual agents (e.g., elderly individuals have a higher probability of contracting a disease) or the relationship between the agents (e.g., transmission rates are higher between parents and small children).   In our model, incarcerated people are considered ``infectious" to those who are most profoundly affected by their absence. Incarcerated people ``transmit" the incarceration ``disease" to their network members with a probability that is a function of the relationship type (e.g. an individual's incarceration has greater effect on his or her child than on a friend) and personal characteristics (e.g. males are more susceptible than females). We denote the probability that infected agent $i$ transmits to susceptible agent $j$ by $p(i \rightarrow j)$.

\begin{figure}[h]
\includegraphics[width = 2.25in]{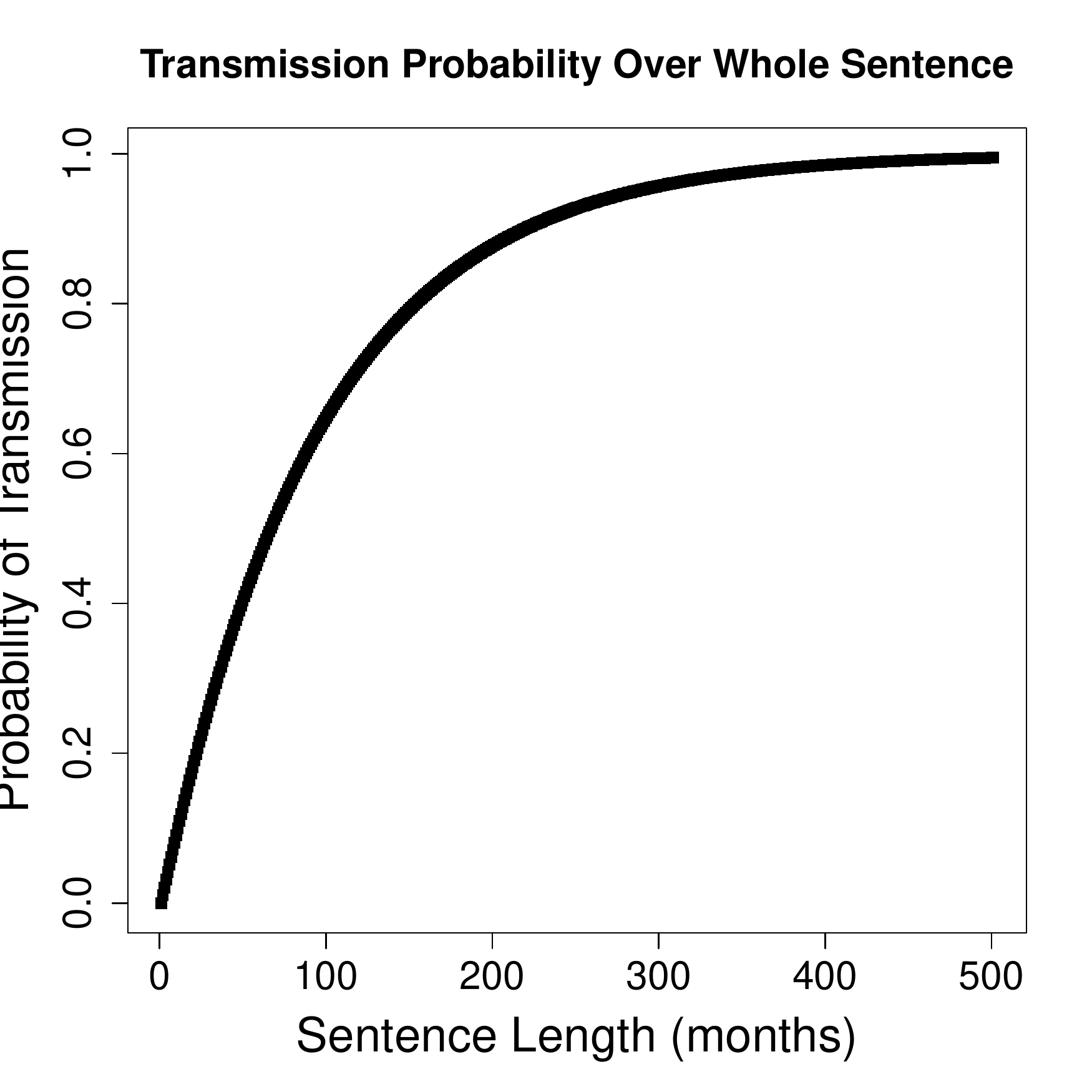}
\caption{\label{fig:functionofsentence} As an example, we assume that at each iteration (month), an agent transmits with probability $p = .1$. This shows the probability that a transmission would have been made throughout the duration of the sentence as a function of sentence length in months. }
\end{figure}

In modeling the spread of disease using the SIS model, the period of infectivity, $s$,  is the duration of time during which the infected individual is contagious. In our model $s$  is the length of the individual's prison sentence, as that is the time during which the inmate's family and close friends are acutely affected. Here, we do not explicitly model increased risk of incarceration due to an inmate's difficulty re-integrating into society. In our model, individuals released from prison cease to be ``infectious" and return to a ``susceptible" state from which they may become re-infected. The source of the new infection will be incarcerated friends and relations, introducing positive feedback into the system. Given the infectious period, the probability that agent $i$ transmits the disease to agent $j$ over the whole course of the infectious period is given by $p_{s}(i \rightarrow j) = 1-(1-p(i \rightarrow j))^s$ i.e. the complement of the probability that agent $i$ does {\it not} transmit to agent $j$ in any of the $s$ iterations during which it is infectious. 
Thus, for fixed $p(i \rightarrow j)$, a longer infectious period results in a higher transmission probability (see Figure \ref{fig:functionofsentence})  and a greater chance that the outbreak becomes widespread. 


\section{Simulation overview}
We synthesize a realistic multi-generational population of agents for which all family and friendship ties are known. All parameters involved in creating this population are based on recent, high quality data. For example, distributions for the sex, lifespan, and the number of children of each agent are taken from the US Census, the Centers for Disease Control and Prevention, and the Social Security Administration, respectively.  The total population consists of 8,856 agents, with 61,376 family and friendship ties. Transmission probabilities are derived directly from the survey of prison inmates presented in \cite{dallaire2007incarcerated}. This survey provides the probability that an inmate's mother, father, sister, brother, or adult child are also incarcerated by inmate gender. That is, $p(i \rightarrow j)$ is taken 
directly from the literature. In our simulation, we treat close friends as siblings in terms of transmission probabilities. We focus on the crime of drug possession, and use data from the Bureau of Justice Statistics to derive sentence lengths by race. 

Using this hypothetical synthetic population, we run ``Black" simulations in which incarcerated agents are assigned sentences that are consistent with those received by Black Americans for drug possession and ``White" simulations using a distribution of sentence lengths corresponding to those received by White Americans for the same crime, as described. {\it In order to test whether differential sentencing alone can explain racial disparities in incarceration rates, we use the same transmission probabilities and the same network to represent both Black and White populations}. 
 
\section{Model Components}
\subsection{Synthetic Population}

We begin with a seed population of $n = 1500$ individuals, $\{a_1, a_2, ..., a_n\}$, from which all members of the population will be descendants. To initialize each agent, it is assigned several attributes.  The $i$th agent, $a_i$,  is assigned a gender from the distribution $g_i \sim \text{Bernoulli}(.5)$, a birth year ($b_i \sim \text{Uniform}(L, U)$), and a spatial location in the unit square ($x_i,y_i \stackrel{iid}{\sim} \text{Uniform}(0,1)$). The location may be thought of geographically, as a physical location in a city, or as a preference space. Regardless of how one prefers to conceptualize the spatial location, it serves the function of creating communities in the network, as friends and spouses are selected with respect to these locations. To simulate a realistic distribution of life durations, the age at death is sampled according to the 2009 life tables released by the Social Security Administration. \footnote{ (\url{http://www.ssa.gov/oact/STATS/table4c6.html})} In these tables, the probability of death in the next year at each age (from 0-119) is given by sex. We treat these as the probabilities of death at each age throughout the simulation. 
If individual $i$ is female and born in year $b_i$, we select a life duration, $l_i$, at random from the distribution given in the female life tables (i.e. $l_i \sim \text{Multinomial}(p_{\text{female}})$). The iteration of death is then given by $d_i=b_i + l_i$. 

Female agents are assigned an age at first birth attribute. Age at first birth is based upon a figure released by the Center for Disease Control, which lists the mean age at first childbirth for women in 2011 as 25.6 years. \footnote{\url{http://www.cdc.gov/nchs/data/nvsr/nvsr62/nvsr62_01.pdf}} 
We specify $h_i$, the age at first birth, to be drawn from the distribution $h_i = 15 + r_i$ where $r_i \sim \text{Poisson}(10.6)$, which has mean 25.6. 


Each female agent is also assigned the number of children she will have throughout her lifetime.
The US Census 
provides the distribution of the number of children women in the age bracket 40-44 have had and  lists the total fertility rate for several years between 1980  and  2008 
, which ranges from about 1.84 to 2.1. Because we do not want our simulated population to die out, 
se adjust the raw distribution given in for 40-44 year old women to be consistent with historical fertility rates. 
Under this adjusted distribution, the expected fertility rate is 2.07 children per woman.

%

\subsection{Network Ties}
In addition to personal characteristics, each agent is endowed with relationships with other agents. These are represented as edges in a network. In our simulation, all parent-child, sibling, spouse and close friend relationships are represented. When an agent reaches its $10$th ``birthday", the agent forms friendship ties. In order to select the number of close friends assigned to each agent, we use data from the 2004 General Social Survey \footnote{\url{http://www3.norc.org/gss+website/}},  
in which respondents indicate the number of individuals with whom they discuss important matters and their relationship to up to five of these people. Because many familial relations are already accounted for in our simulation, we count only those people listed who are not parents, children, spouses or siblings to calculate the probability of selecting each possible number of friends, $f_i$. 


Conditional on $f_i$, we  select the specific agents that will be designated as friends. We consider potential friends to be non-siblings between the ages of 9 and 11. This age range is based on information obtained from the National Longitudinal Study of Adolescent Health \footnote{\url{http://www.cpc.unc.edu/projects/addhealth}; data not publicly available} 
, in which children were asked to list several of their friends. Of the friends listed, nearly 95\% were within one grade of the student (75\% shared the same grade). Because friends within one year of the students made up the majority of close childhood friendships, we also restrict to this range. From this set of potential friends, we select the $f_i$ agents that are closest in Euclidean space to the given agent.  
 

 In the iteration in which a female actor's first child is born, she is assigned a partner. In this hypothetical population, we only model opposite-sex spousal ties. The algorithm first finds all potential partners -- unrelated and non-friend male agents whose current age is between the female agent's age and nine years older. We selected this scheme based upon data from the 1999 US Census that shows that for about 80\% of marriages, the husband's age minus the wife's age falls into the range $[-1, 9]$. We restrict male agents to be strictly older than female agents, as our partner selection algorithm tends to force the age of the male partners to fall to the lower end of the allowable range. From this collection of potential partners, the agent is assigned the potential partner that is located the closest to it, again using Euclidean distance. 

  At this point, the children of the couple (or single mother) are initialized. The first child is assigned to be born in the current iteration. Subsequent children are assigned birth years as the current iteration plus independent and identically distributed draws from a Poisson distribution with mean parameter $\lambda =4.5$. The mean parameter, $\lambda$, is again based on the Centers for Disease Control's data used to calibrate the age at first birth. 
  Child locations are set to be half way between the mother and father's location plus random noise ($\text{Uniform}(m_1 - .05, m_2 + .05)$), where $m_1$ and $m_2$ are the midpoints between the mother and father's location along the $x$ and $y$ axis, respectively.

\begin{figure}[h]
\includegraphics[height = 1.65in]{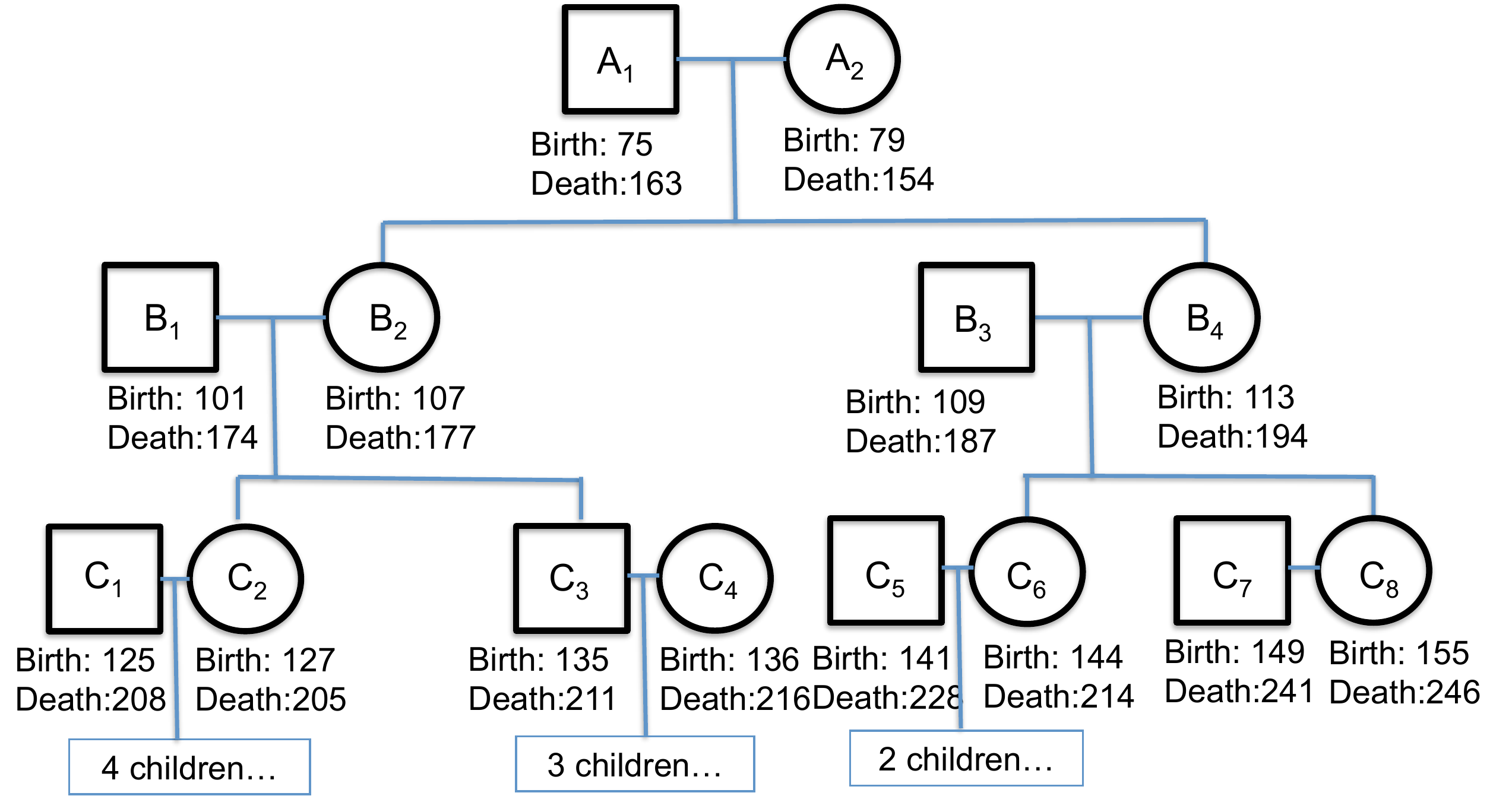}
\includegraphics[height = 1.65in]{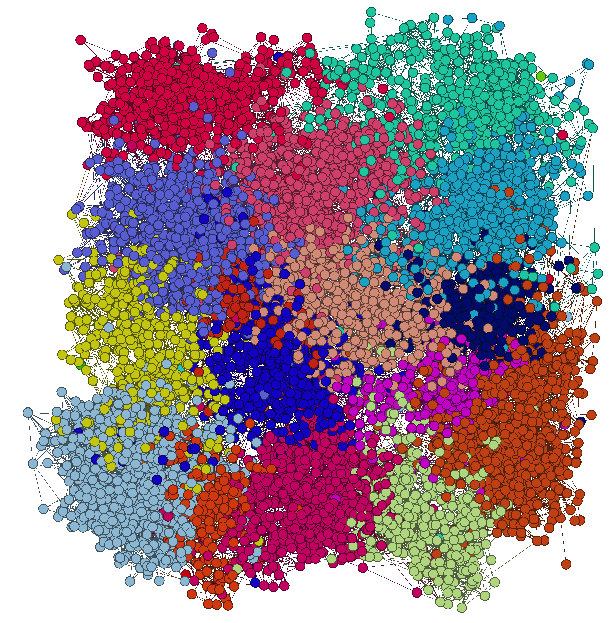}

\caption{\label{fig:familytree} (left)  One family from the synthetic population (right) The network of agents in our simulation. Different colors indicate communities found in the network. These communities are not explicitly included in the simulation}
\end{figure}

  We run our algorithm for 200 iterations, resulting in a total population of 13,826 individuals. We discard the first 150 iterations as a burn-in period, reducing dependence on our initial conditions. Those agents that are part of our simulation (i.e.\ those that are alive at any point beyond the 150th iteration) are retained, resulting in a population of size 8,856. The left panel of Figure \ref{fig:familytree} shows an example family tree generated by our algorithm. This individual family, of course, does not exist in isolation of the rest of the population. The population-wide network is shown in the right panel. 

%
%

%
 
\section{Generating Sentences}
\label{sec:sentences}
\begin{figure}[h]
 \centering
\includegraphics[width = 3in]{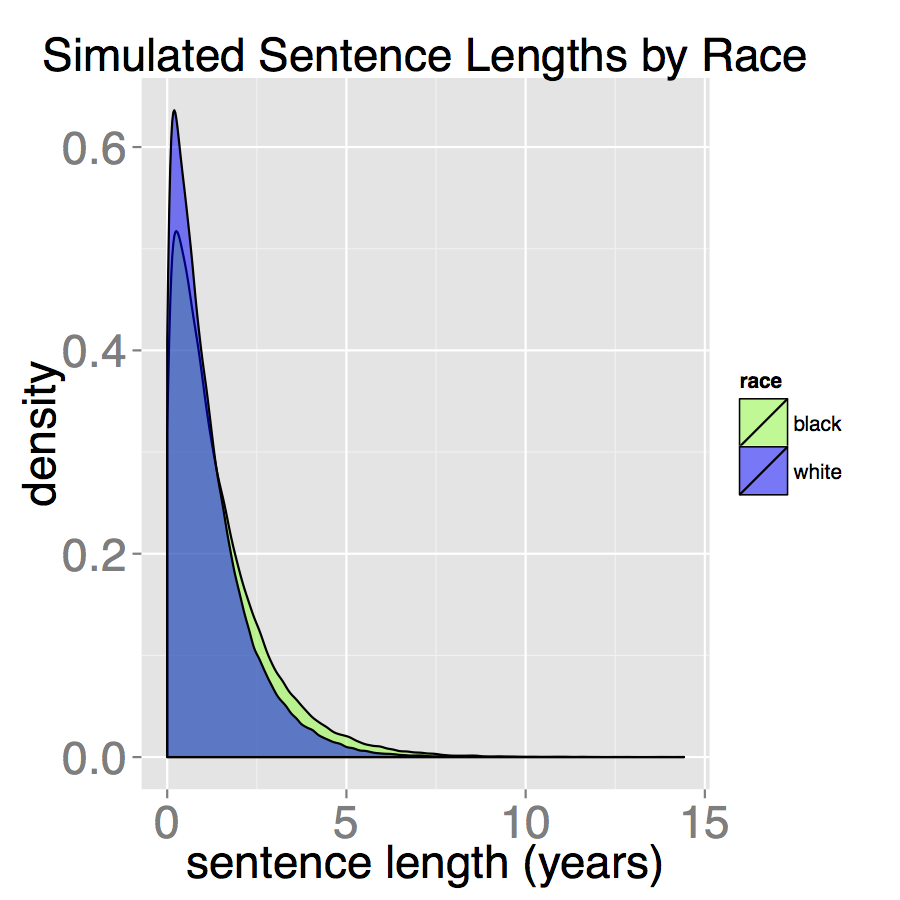}
\caption{\label{fig:sentencelengths}Comparison of the distribution of sentence lengths for Whites and Blacks under our simulation parameters. These distributions are fitted to data released by the Bureau of Justice Statistics. }
\end{figure}

Data released by the Bureau of Justice Statistics\footnote{\url{http://www.bjs.gov/index.cfm?ty=pbdetail\&iid=2056}}
indicates that the duration of sentence served for the same crime varies by race. In particular, the Bureau of Justice Statistics states that for drug possession, the mean sentence for Whites is 14 months with a median of 10 months. For Blacks, the mean sentence served is 17 months with a median of 12 months. We use a negative binomial distribution to generate sentence lengths that are consistent with the specified summary statistics.  A comparison of the sentence distributions is shown in Figure \ref{fig:sentencelengths}.


\section{Transmission Probabilities}
\label{sec:probabilities}
Dallaire \cite{dallaire2007incarcerated} presents the results of a survey of incarcerated individuals. In this survey, each inmate is asked which of their relations are also incarcerated. The proportion of inmates whose relations are incarcerated are reported in this data by sex. We use these probabilities to derive our transmission probabilities, $p(i \rightarrow j)$. We have noted that if  agent $i$ has probability $p(i \rightarrow j)$ of infecting agent $j$ each month it is incarcerated and its sentence is $s$ months, then the probability of transmission over the course of its incarceration, $p_{\text{sentence}}(i \rightarrow j)$, is given by,  $p_{\text{sentence}}(i \rightarrow j) = 1-(1-p(i \rightarrow j))^s$.  One can easily solve for $p(i \rightarrow j) = 1-(1-p_{\text{sentence}}(i \rightarrow j))^{\frac{1}{s}}$. We set $s=14$ to approximately calibrate to a value between the Blacks and Whites.  The derived monthly transmission probabilities used in this simulation are given in Table \ref{tab:transprobs}. 


\begin{table}[ht]
\centering
\caption{\label{tab:transprobs}{\small Derived monthly transmission probabilities, $p(i \rightarrow j)$. }}
\begin{tabular}{r|c|c}
  \hline
 & ~women~   & ~~men~~   \\ 
  \hline
mother ~& 0.001 & 0.003 \\ 
  father ~& 0.011 & 0.011 \\ 
  sister ~& 0.008 & 0.004 \\ 
  brother~ & 0.033 & 0.030 \\ 
  spouse ~& 0.004 & 0.001 \\ 
  adult child ~& 0.017 & 0.006 \\    \hline
\end{tabular}
\end{table}
The derived monthly transmission probabilities are most usefully understood in the context of the probability of transmission averaging over sentence length, i.e. the marginal transmission probabilities,

\begin{equation}
p_{\text{race}}(i \rightarrow j) = \sum_{s = 0}^{\infty} p_{\text{sentence}}(i \rightarrow j) \pi_{\text{race}}(s),
\end{equation}
where $p_{\text{sentence}}$ is defined above 
and $\pi_{\text{race}}(s)$ is the distribution of sentence lengths for each race.  Marginal probabilities by race, sex of inmate, and relation are given in Table \ref{tab:margprobs}, along with the original probabilities listed in \cite{dallaire2007incarcerated}.  These were calculated using the Monte Carlo method. We notice that our marginal probabilities for the White sentences tend to be just slightly lower than those given. For Black sentences, the marginal probability of transmission tends to be slightly higher than that given. This is, of course, unsurprising as the sentences tend to be shorter for Whites and they thus have fewer opportunities for transmission. Recall that the monthly transmission rates under the two scenarios are precisely the same, so one should not interpret the differences in marginal probabilities to mean that this model implies that Blacks are more susceptible than Whites under the same conditions. The only differences that exist here are due to the discrepancies in sentencing. We find that the probability of transmission for Blacks under these parameters tends to be about 20\% greater than the probability of transmission for Whites. 

\begin{table*}
\centering
\caption{\label{tab:margprobs}{\small Probabilities given in \cite{dallaire2007incarcerated} and marginal transmission probabilities for Whites and Blacks.}}
\begin{tabular}{r|c|c|c|c|c|c}
 & \multicolumn{2}{c}{Survey\hspace{5mm}} & \multicolumn{2}{|c|}{White\hspace{5mm}} & \multicolumn{2}{|c}{Black} \\
  \hline
 & ~women~& ~~men~~     & ~women~ & ~~men~~ & ~women~ & ~~men~~  \\ 
  \hline
mother ~& 0.012 & 0.048 & 0.012 & 0.046 & 0.014 & 0.056  \\ 
  father ~& 0.147 & 0.148 & 0.138 & 0.138 & 0.163 & 0.163  \\ 
  sister ~& 0.107 & 0.059 & 0.101 & 0.058 & 0.121 & 0.069  \\ 
  brother ~& 0.377 & 0.349 & 0.324 & 0.303 & 0.370 & 0.347  \\ 
  spouse ~& 0.059 & 0.011 & 0.057 & 0.011 & 0.069 & 0.013  \\ 
  adult child ~& 0.213 & 0.085 & 0.194 & 0.082 & 0.227 & 0.098 \\ 
     \hline
\end{tabular}
\end{table*}

\section*{Results} 
We run our simulation 250 times using each sentence length distribution, resulting in 250 ``Black" epidemics and 250 ``White" epidemics.  The populations are initialized to have approximately 1\% of individuals incarcerated at the beginning of the simulation. Although White and Black incarceration rates, in reality, were different in the mid-80s, we initialize the simulations equivalently under the two scenarios to rule out the possibility that the resultant disparities are due to initial conditions alone. Our first analysis looks at the effects of differential sentencing after 50 years. Figure 2a shows the mean epidemic curves and corresponding confidence intervals by race.  Figure 2b shows several example trajectories. While the prison epidemic takes off 
under Black sentence lengths, reaching just under 3\% incarcerated on average after 50 years, 
the model using White sentence lengths indicates that the prison population first declines and then increases at a much slower rate, reaching about 0.725\% over the same time period. Figure 2c shows the results of removing transmission from the model, i.e.\ simulating a situation in which agents spontaneously become incarcerated independent of any relationship with an incarcerated person. We set the probability of spontaneous infection such that at the end of the simulation, the Black population's prevalence is roughly 3\% as in the SIS simulations.  Based on this, we conclude that sentence length disparities in the absence of network effects do not account for the observed difference in incarceration rate, as this model results in a difference between the two curves of only about 20\%. It is feedback through the local network coupled with sentencing disparities that causes such large differences in incarceration rates.

\begin{figure}[h]
\includegraphics[width = 6in]{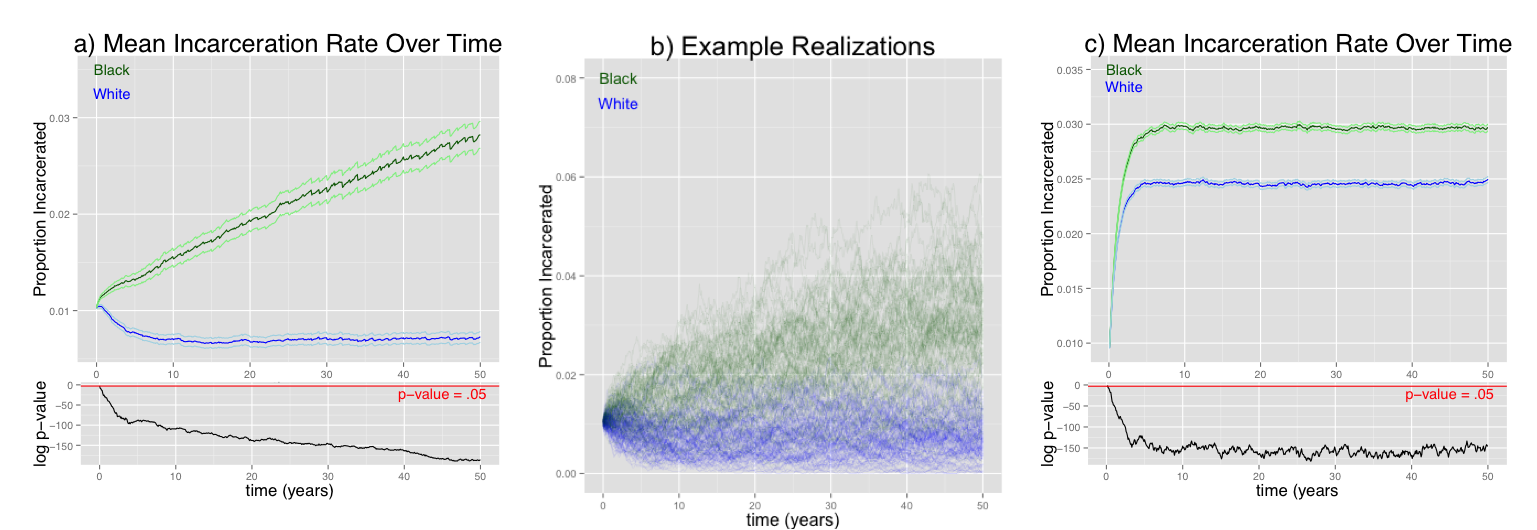}
\caption{(a) The mean incarceration prevalence by race. Log p-values shown below indicate that at all but the first time point, the mean prevalence is significantly different between the two populations. (b) Several example epidemic curves. (c)The mean incarceration prevalence under a non-contagious model.}
\end{figure}

\begin{figure}[h]
\centering
\includegraphics[width = 3.5in]{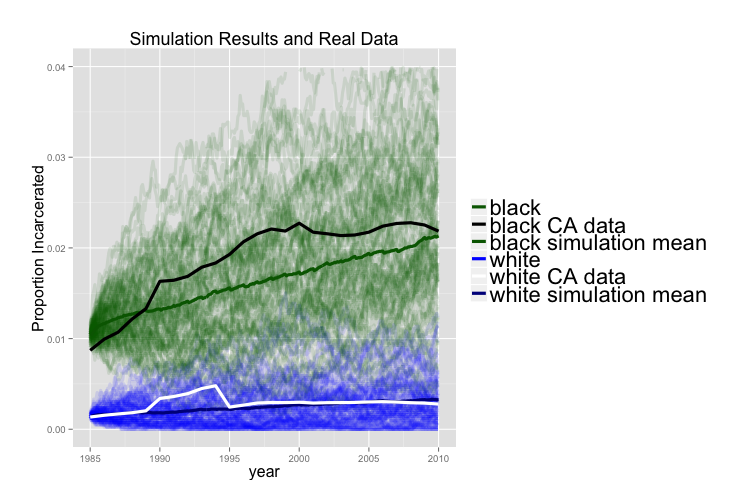}
\caption{The incarceration rate in California by year and race. The number of incarcerated people by race and year was calculated using the National Prisoner Statistics data set from the Inter-University Consortium for Political and Social Research. The total number of people in California by race and year was calculated from the California Department of Finance.}
\label{fig:ca-epi}
\end{figure}

Finally, as a comparison with real data, we initialize the Black and White simulations to be near the real incarceration rates in California in the mid-1980s when mandatory drug sentencing became law (1\% and 0.15\% for Blacks and Whites respectively). The results of this are shown in Figure \ref{fig:ca-epi}. From 1986 to 2010, the incarceration rate for Blacks in California climbed from about 1\% to 2.18\%, whereas the rate for Whites rose much more modestly from 0.15\% to 0.277\%. In our simulations, the Black and White simulations increase to 2.25\% and 0.34\%, respectively, over the same time period. For the black trajectory, our mean trajectory deviates slightly though the ultimate results after 25 years are remarkably similar to reality; however, we note that the real trajectory in California is well within the range of values that are typical under our simulation, as it falls well within the cloud of trajectories shown. The white trajectory in California is quite similar to the mean trajectory in our simulation. 

\begin{figure}[h]
\centering
\includegraphics[width = 6in]{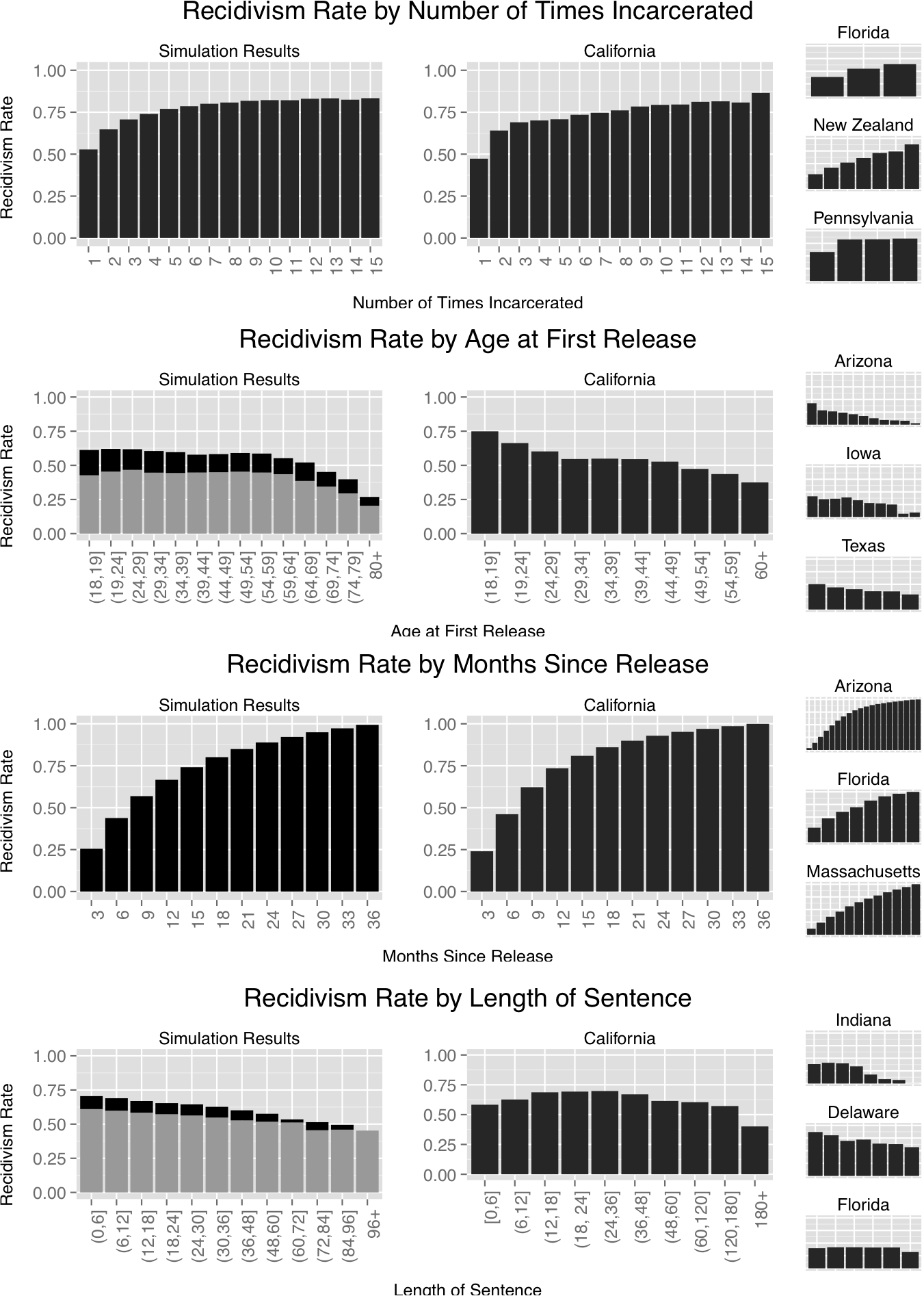}\\
\caption{\label{fig:recidivism}A comparison of the output of our simulation model with recidivism rates in California and several other states. }
\end{figure}

Additionally, in Figure \ref{fig:recidivism} we compare the recidivism rates derived from the simulation model to those released by the state of California. The plots to the right show the same statistics from a variety of other states. These vary because not all states release the same statistics; we chose California as our primary point of comparison because its report shows recidivism rates by the most factors and at the highest level of discretization. Remarkably, our model reproduces the structural properties of recidivism very accurately. For example, in both our model and the data from California, recidivism rates increase with the number of times an inmate has been incarcerated, with the largest increase occurring between the first and second incarceration. This effect emerges as a structural property of the SIS model {\it without including an increased probability of incarceration for those who have previously been incarcerated}. An agent's incarceration affects its local network enough that, upon release, it has a higher probability of return. In all cases, including our simulation results, the recidivism rate is lower for those who are released at an advanced age. In fact, our simulation even reproduces a subtle demographic bulge in rates that is apparent in the California report. This is again an emergent property of the model -- as an agent ages, it tends to have fewer (and different types of) contacts who generate positive feedback for incarceration. The rate by months since release (for those who recidivated within three years) closely matches the structure of the data from all states. These results suggest that our model is reflecting the underlying regularities in the system.

\subsection{An ordinary differential equations (ODE) approach}
Under assumptions of random mixing and homogeneity of transmission rate, the SIS model can be written as a set of ordinary differential equations. In this case, the number of people infected -- the {\em prevalence} -- when an outbreak reaches a steady state, $I$, is determined by the expected number of transmissions per infected person, which in turn is given by the product $ps$ of transmission rate and the duration of infectivity. The presence of a positive feedback loop makes the relation between prevalence and number of transmissions highly nonlinear. In particular, ignoring births and deaths, $I = 0$ if $s < p^{-1}$ and $1 - 1 /ps$ otherwise. Thus, given the transmission rate $p$, we can define a critical duration of infectivity $s_c \equiv p^{-1}$ such that for $s < s_c$ the outbreak dies out, while for $s>s_c$ it achieves a non-zero steady-state prevalence.  Near $s_c$, small differences in sentencing (i.e.\ duration of infectivity) can cause large differences in incarceration (i.e.\ disease) prevalence.

We take an agent-based simulation approach to modeling the incarceration epidemic because neither the assumption of uniform mixing nor the assumption of homogeneity are met.
Indeed, the network of family and friends plays a {\em crucial} role in our hypothesis. Furthermore, the data show that transmission rates depend on the nature of the relationship between the infectious and susceptible people, and any particular susceptible may simultaneously have several types of relationships with different infectious people (e.g.\ mother {\em and} sister {\em and} daughter). It is easier to capture this heterogeneity in transmission rates in an agent-based simulation than in a set of ODEs.

 Moreover, using the output of our simulation, we can \emph{generate} a population-wide mean transmission rate $p$ to calibrate an ODE model. The result is $p \approx 0.0612$ transmissions per infected person per month, or $s_c \approx 16.3$ months. Thus the mean sentence lengths (17 and 14) are on opposite sides of the critical point. Under this model,  the incarceration epidemic in the White population would eventually die out, as no spontaneous infections are allowed. However, the Black population's incarceration rate would reach a steady state of about 3.9\%. As this model does not account for spontaneous infections, this result is consistent with those of the agent-based model. This approach, however, does not allow us to assess the validity of the model or comment on its ability to reproduce structural properties of the epidemic using other withheld sources of information, such as the recidivism data we have shown. 


\section{Discussion}

The model presented here demonstrates that the dramatic disparities in incarceration rates of Black and White Americans can be explained by the ``transmission" of incarceration from an incarcerated person to his or her close associates combined with modest differences in sentencing.  A relatively small difference in sentencing of, on average, three months created incarceration discrepancies similar to those observed today. Further evidence supporting the plausibility of the model was found in its agreement with observed patterns of recidivism, especially in California. However, the model reveals that, contrary to the arguments of some advocates, sentencing differences alone do not account for the disparities. To generate the vast incarceration disparities observed today in the U.S., the model must include both sentencing disparities and a mechanism for transmitting incarceration similar to that in the SIS model of disease propagation. 

Our model is silent on the question, ``Why is there a disparity in sentencing?''
However, it demonstrates that disparities in incarceration rates between White and Black Americans may have as much to do with the social construction of crime as they do the criminal behavior of individuals.


\section*{acknowledgments}
The authors thank the Network Dynamics and Simulation Science Laboratory team, the SAMSI ABM working group,  and Jose Torres for their valuable input. This work has been partially supported by DTRA Grant 
HDTRA1-11-1-0016 and the National Institute of General Medical Sciences of the National Institutes of Health (NIH) under the Models of Infectious Disease Agent Study (MIDAS) program, award number 2U01GM070694-09. The content is solely the responsibility of the authors and does not necessarily represent the official views of the NIH or DTRA.

\end{document}